\documentclass{article}

\usepackage{arxiv}

\usepackage[utf8]{inputenc} 
\usepackage[T1]{fontenc}    
\usepackage{hyperref}       
\usepackage{url}            
\usepackage{booktabs}       
\usepackage{amsfonts}       
\usepackage{nicefrac}       
\usepackage{microtype}      
\usepackage{lipsum}		
\usepackage{graphicx}
\usepackage{gensymb}
\usepackage{doi}
\usepackage[numbers]{natbib}

\title{A Novel Method to Eliminate the Symmetry Dependence of Fiber Coils for Shupe Mitigation}

\author{
Tugba Andac Senol\\
  Department of Physics\\
  Nanotechnology Research Center (NANOTAM) \\
  Bilkent University \\
  Ankara\\
  \texttt{\ tugba.andac@bilkent.edu.tr*} \\
   \And
  Onder Akcaalan\\
  Nanotechnology Research Center (NANOTAM) \\
  Bilkent University \\
  Ankara\\
  \texttt{\ oender.akcaalan@desy.de} \\
  \And
  Aylin Yertutanol\\
  Nanotechnology Research Center (NANOTAM) \\
  Bilkent University \\
  Ankara\\
  \texttt{\ karagoz@bilkent.edu.tr} \\
  \And
   Ekmel Ozbay\\
  Department of Physics\\
  Nanotechnology Research Center (NANOTAM) \\
  Department of Electrical and Electronics Engineering\\
  Institute of Materials Science and Nanotechnology\\
  Bilkent University \\
  Ankara\\
  \texttt{\ ozbay@bilkent.edu.tr} \\
}

\renewcommand{\shorttitle}{\textit{arXiv} Template}

\hypersetup{
pdftitle={A template for the arxiv style},
pdfsubject={q-bio.NC, q-bio.QM},
pdfauthor={David S.~Hippocampus, Elias D.~Striatum},
pdfkeywords={First keyword, Second keyword, More},
}

\begin{document}
\maketitle

\begin{abstract}
	It is a well-known fact that interferometric fiber optic gyroscopes (IFOGs) are easily distorted by thermal effects and distortion results in the degradation of the performance of these sensors. Changing the fiber coil geometry, increasing the winding symmetry, adding fiber buffer layers around the fiber coil, using different modulation methods for multifunctional integrated optic chips (MIOCs), and using special types of fibers, such as photonic crystal fibers (PCFs), are some alternative solutions for preventing this degradation. This paper, theoretically and experimentally, investigates not only how different types of fiber coil winding methods behave under different rates of temperature change but also presents a novel method, to the best of our knowledge, to eliminate the Shupe effect, without violating the simplest IFOG scheme. This method rules out the importance of the winding symmetry epochally and the need of any extra treatment for the fiber coil to increase the thermal performance of the system. Regardless of the symmetry of the fiber coil winding, the rate error due to the Shupe effect can be reduced to about $\pm$$0.05^\circ/$h for any rate of temperature change with this new method according to the experimental results.
\end{abstract}


\section{Introduction}
Applications, such as guidance, navigation, and control systems in air and land vehicles, both in industry and military fields, require compact, cost-effective, and reliable inertial navigation systems equipped with a highly sensitive gyroscope that uses the Sagnac Effect \cite{sagnac1913ether}. The effect that Sagnac unearthed led to the advent of IFOG with improvements of low-loss optical fibers, as well as solid-state semiconductor light sources and detectors \cite{lefevre2022fiber}. Moreover, the fact that IFOGs are solid-state devices and have no moving parts, the ease of achieving higher sensitivity by increasing the number of wraps in the fiber coil used \cite{lefevre2014fiber}, their tendency for miniature manufacturing \cite{nayak2011fiber}, long lifetime \cite{yin2017fiber}, very quick turn-on time \cite{yin2017fiber}, high reliability \cite{lefevre2014fiber, lefevre2013fiber, korkishko2012interferometric} and high precision are just some of the advantages that have ensured their outdistancing the other types of gyroscopes, e.g., ring laser gyroscopes (RLGs) \cite{yin2017fiber, korkishko2012interferometric, lefere1997fundamentals, lefevre2012fiber}.

\begin{figure}[!t]
\centering
\includegraphics[width=0.8\textwidth]{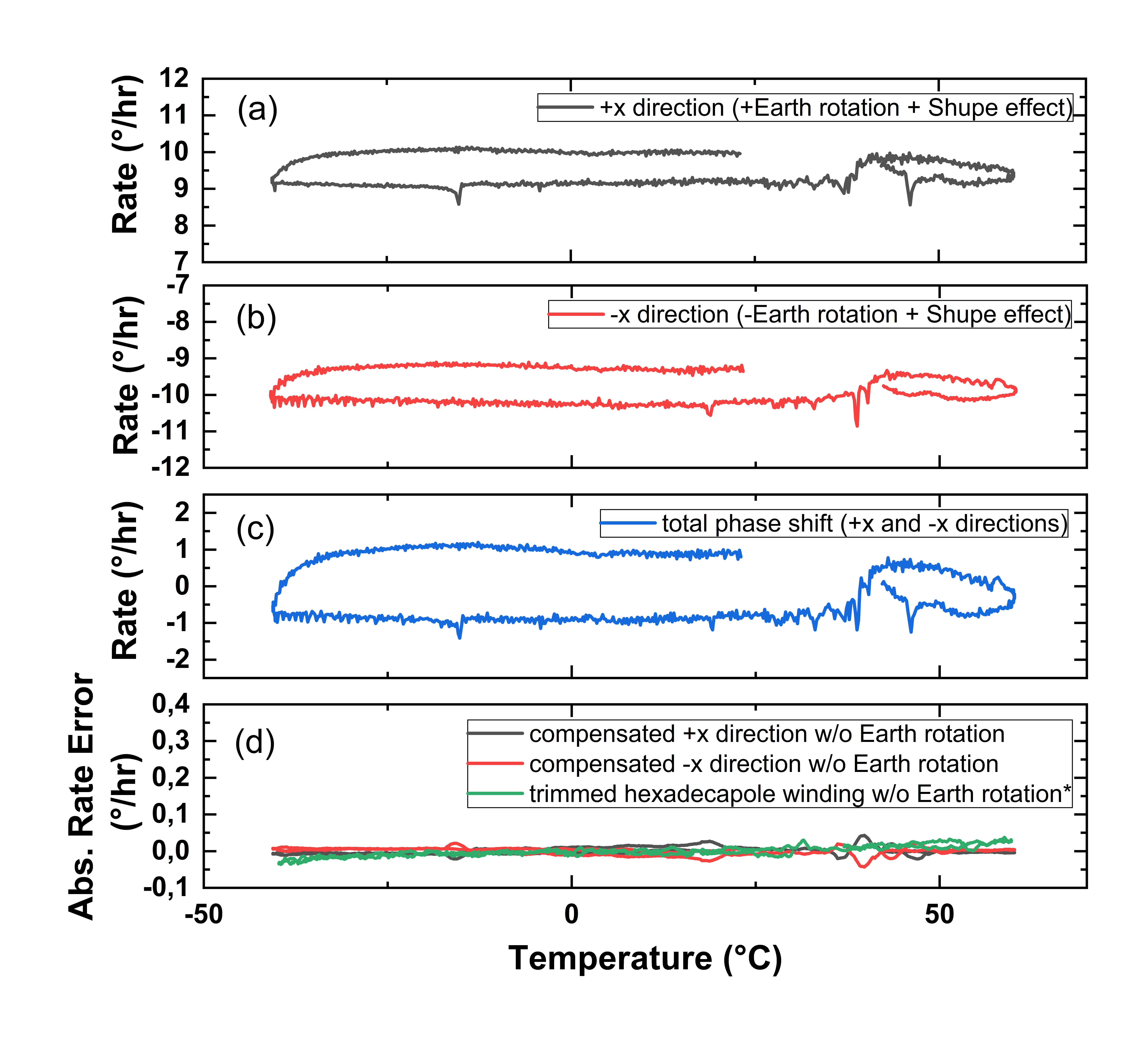}
\caption{\label{fig:Fig1} Rate vs. temperature graph of the IFOG comprising the fiber coil with the hexadecapole pattern obtained at a rate of 0.2\degree C/min temperature change horizontally placed in the (a) +x direction, (b) horizontally reversed placed in the -x direction, (c) total phase shift of the same IFOG in both directions, (d) compensated absolute rate errors in both directions compared with the result obtained from the IFOG comprising a perfectly trimmed hexadecapolar fiber coil at a rate of 0.2\degree C/min temperature change.}
\end{figure}

An ideal IFOG should be able to only measure the Sagnac phase shift. However, the Sagnac effect is not the only source of measurable phase shift in practice. Environmental effects such as temperature \cite{shupe1980thermally}, magnetic field \cite{bohm1982sensitivity, yertutanol2021fiber}, or vibration \cite{ohno1992development} create differences in the optical paths and can cause some undesired nonreciprocal phases apart from the Sagnac phase shift. Because of the difficulty of distinguishing the real Sagnac phase shift from the other nonreciprocal phase shifts, reducing these nonreciprocal errors caused by the environmental effects becomes a necessity and the success of this reduction eventually determines the quality of these fiber optic rotation sensors.

As the sensing element, the fiber coil takes the lead in affecting its performance. In particular, the thermal sensitivity of the fiber coil reveals the sensitivity of the IFOG. Special winding methods have been proposed to overcome this performance limitation \cite{frigo1983compensation, mile2009octupole, chomat1993efficient, wang2018design}. The quadrupole winding pattern is one of the most widely used techniques with proven performance \cite{frigo1983compensation}. It helps to reduce the non-reciprocity caused by the false rotation signal when the counter-propagating light waves do not follow the same paths through the fiber segments in the fiber coil exposed to a different rate of temperature changes. However, since this winding method is no longer sufficient to meet the performance demands of IFOGs, further work is required to increase the winding symmetry of the fiber coil and eventually improve the thermal performance of the IFOG \cite{hong2023accurate}. Apart from the winding methods, mechanical studies \cite{zhang2020method} and different modulation techniques \cite{he2019stability, chen2023suppression} have also been carried out to reduce the temperature effects. Using special types of fibers such as photonic crystal fibers (PCFs) in fiber coils to decrease the temperature dependency of the refractive index is also offered as an alternative to solve the problems caused by the environmental effects on IFOGs \cite{song2023advanced}. 

In the present paper, we study the rate errors due to the thermal characteristics of different types of fiber coils both theoretically and experimentally. We bring in the trimming method in addition to the usage of different winding methods for the very same purpose. We verify the model experimentally as well with different rates of temperature change. Furthermore, in noticing the rate error due to the Shupe effect behaving independently in the direction of the real rotation during the temperature analyses, we developed a novel method in which the Shupe effect can be easily measured and integrated into the IFOG systems independent of temperature variation and fiber coil winding symmetry. 
\section{Methods and Analysis}
Beams traveling uneven paths create undesirable nonreciprocal phase shifts along with the real phase shift coming from the rotation that is to be measured and by considering this phase shift, the angular error of thermally induced non-reciprocity, $\phi_{shupe} (t)$, for the IFOG can be calculated by \cite{shupe1980thermally}

\begin{eqnarray}
\label{eqnarray:1}
\phi_{shupe} (t) = \frac{n_c}{4NA}(\frac{dn_c}{dT}+\alpha n_c)(\int_{0}^{L/2} dl(2l-L)\times[\Delta T (l,t) - \Delta T (l',t)]),
\end{eqnarray}

where $n_c$ = $1.46$ is the refractive index of the fiber core, $N = 48$ is the number of turns in the fiber coil, $A$ is the area of the fiber coil with a diameter of $10$ cm, $dn_c/dT$ = $10^{-5}$$ / ^{\circ}C$ is the temperature dependence of the refractive index of the fiber, $\alpha = 5x10^{-7} / ^{\circ}C$ is the coefficient of linear thermal expansion of fiberglass and $L$ = $1037$ m is the total length of the fiber coil. Here, $l$ and $l'$ represent the locations of the two rotating beams at the same time and $\Delta T(l,t)$ and $\Delta T(l',t)$ represent the temperature differences between the locations $l$ and $l'$ simultaneously.

The simplified version of the equations, as the simulated and experimental works of thermal effects for the winding methods of four different fiber coils (dipole (AB), quadrupole (ABBA), octupole (ABBABAAB), and hexadecapole (ABBABAABBAABABBA)), are presented in the supplementary material.

\begin{figure}[!b]
\centering
\includegraphics[width=0.8\textwidth]{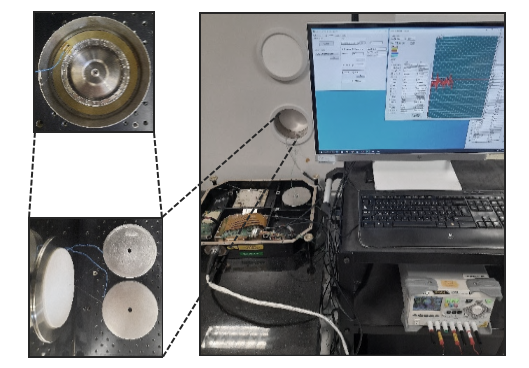}
\caption{\label{fig:setup_photograph_3}Photograph of the experimental setup}
\end{figure}

As seen in Supp. Eqs. (2) and (3), the rate error due to the Shupe effect is an absolute error that is independent of the direction of the real rotation. The phenomenon of the Shupe effect being independent of real rotation allows us to easily calculate the rate error due to this effect by simply reversing the fiber coil axis horizontally, such as changing its axis from the +x to the -x. By considering a constant rotation such as the Earth's rotation applied to the system; the total phase shifts for the +x and -x axes can be shown as in Eqs. (\ref{eqnarray:4}) and (\ref{eqnarray:5}) ,respectively. When these phase shifts are added up to each other as shown in Eqn. (\ref{eqnarray:6}), the twofold pure rate error due to the Shupe effect can be easily found.

\begin{eqnarray}
\centering
&&\label{eqnarray:4}\phi_{total}^{+x} =  \phi_{rotation} + \phi_{shupe_1}\\
&&\label{eqnarray:5}\phi_{total}^{-x} = -\phi_{rotation} + \phi_{shupe_1}\\
&&\label{eqnarray:6}\phi_{total}^{+x} +\phi_{total}^{-x} = 2\phi_{shupe_1}
\end{eqnarray}

\begin{figure}[!b]
\centering
\includegraphics[width=0.8\textwidth]{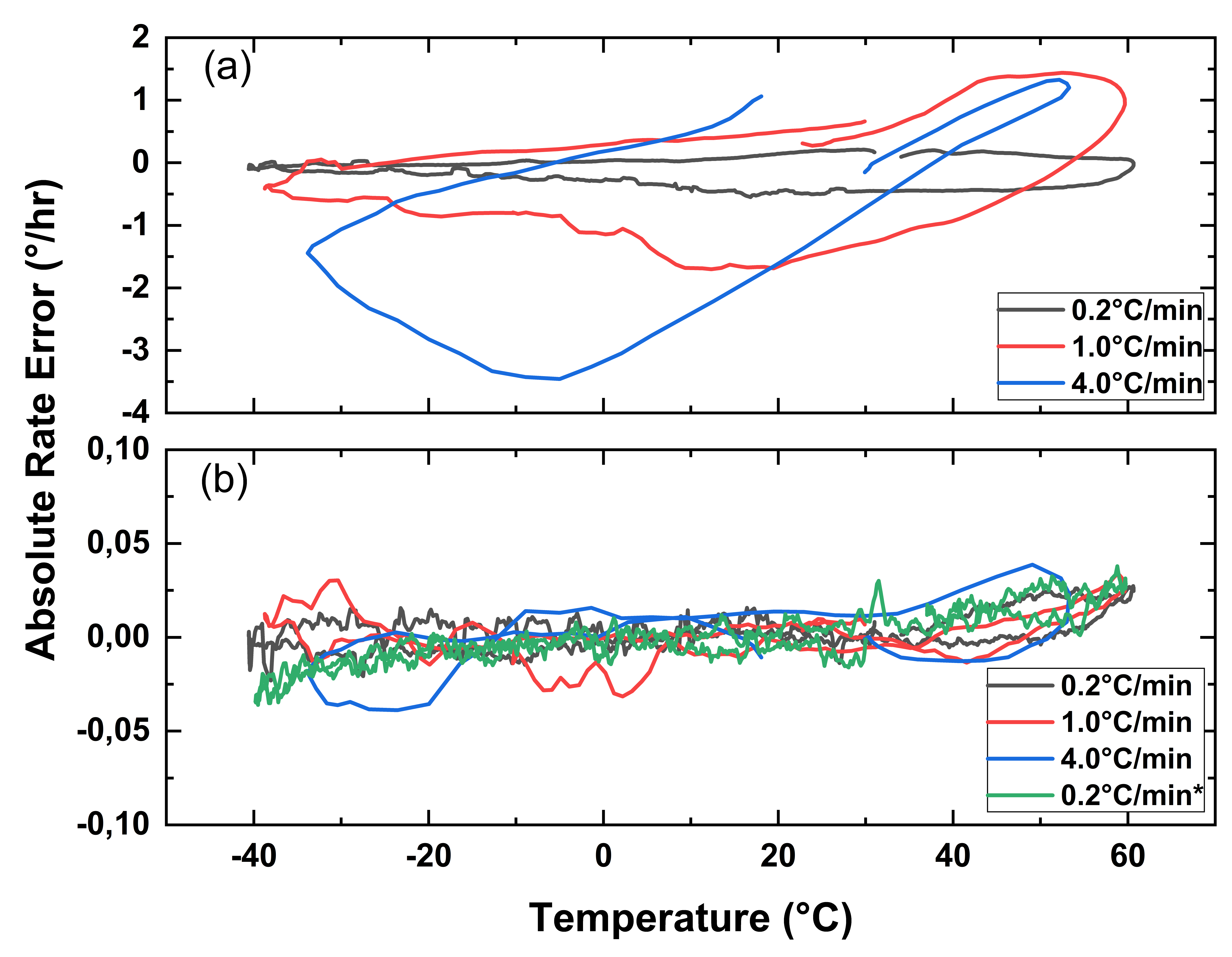}
\caption{\label{fig:Figure_3} (a) Absolute rate errors of the IFOG comprising the hexadecapolar fiber coil obtained at a rate of 0.2\degree C/min., 1\degree C/min., and 4\degree C/min. temperature change, (b) compensated absolute rate errors of the hexadecapolar fiber coil, respectively, compared with the result obtained from the IFOG comprising a perfectly trimmed hexadecapolar fiber coil at a rate of a 0.2\degree C/min. temperature change.}
\end{figure}

To prove this phenomenon experimentally, we used a hexadecapolar fiber coil wound by using PM fiber with 80 $\mu$m cladding and 168 $\mu$m coating diameter on a spool with an average diameter of 85 mm consisting of 48 layers in total and each layer had 80 turns of fiber as was also used in the model (see supplementary material). The total length of the fiber coil was approximately 1037 m and was obtained by an optical time domain reflectometer (OTDR) measurement (YOKOGAWA AQ7270). We constructed an IFOG by splicing a three-in-one homemade multifunctional integrated optic chip (MIOC) fabricated with annealed proton exchange (APE) method to the fiber coil that was also spliced to a homemade broadband ASE light source with a central wavelength of 1537 nm coupled to a 3-dB optical coupler. APE-MIOC was acting as a $>$40-dB polarizer, a 3-dB coupler and a phase modulator. Once the loop was completed, we implemented a photodetector to the system to convert the interfered optical signal. We packed and placed only the fiber coil with a temperature sensor in a magnetic shield to prevent nonreciprocal errors coming from magnetic field changes. First, we ran the test while the fiber coil was horizontally placed in the +x axis over different temperatures in the climatic chamber. Then, we horizontally reversed the fiber coil and repeated the same test. The rotation rate of Earth was measured as +9.6\degree /h in Ankara, Turkey via the constructed IFOG with horizontally placed fiber coil and -9.6\degree /h via the constructed IFOG with horizontally reversed placed fiber coil. The rate data were read with the aid of software by using a closed-loop modulation technique and presented as a function of temperature assessed between -40\degree C and +60\degree C operating temperature in the chamber with a 0.2\degree C/min. rate of temperature change.

Figure~\ref{fig:Fig1} shows the experimental rate vs. temperature data obtained from the IFOG comprising the fiber coil placed horizontally in the +x direction (a), horizontally reversed in the -x direction (b) and the total phase shift in both directions (c) calculated by using Eqn. (\ref{eqnarray:6}) from both tests respectively. To compare the rate errors due to the Shupe effect, the Earth rotation is subtracted from the real rates, and, therefore, the compensated absolute rate errors that are reduced to $\pm$$0.05^\circ/$h are presented with the result obtained from the IFOG comprising a perfectly trimmed hexadecapolar fiber coil (d). The comparison between these three results shows good agreement with each other.

In IFOG systems, it is not possible to simultaneously reverse the fiber coil axis horizontally. Therefore, a separate setup has been prepared to see if the presented phenomenon will still be valid when two different fiber coils, one of which is horizontally reversed, are used in the setup, as shown in Figure~\ref{fig:setup_photograph_3} and similarly as explained herein above. Then, the equations turn to:

\begin{eqnarray}
\centering
&&\label{eqnarray:7}\phi_{totalC1}^{+x} =  \phi_{rotation} + \phi_{shupe_{C1}}\\
&&\label{eqnarray:8}\phi_{totalC2}^{-x} = -\phi_{rotation} + \phi_{shupe_{C2}}\\
&&\label{eqnarray:9} R_{C1,C2} = \phi_{shupe_{C1}}/\phi_{shupe_{C2}}\\
&&\label{eqnarray:10}\phi_{totalC1}^{+x} +\phi_{totalC2}^{-x} = \phi_{shupe_{C1}} + \phi_{shupe_{C2}}
\end{eqnarray}

\begin{figure}[!b]
\centering
\includegraphics[width=0.8\textwidth]{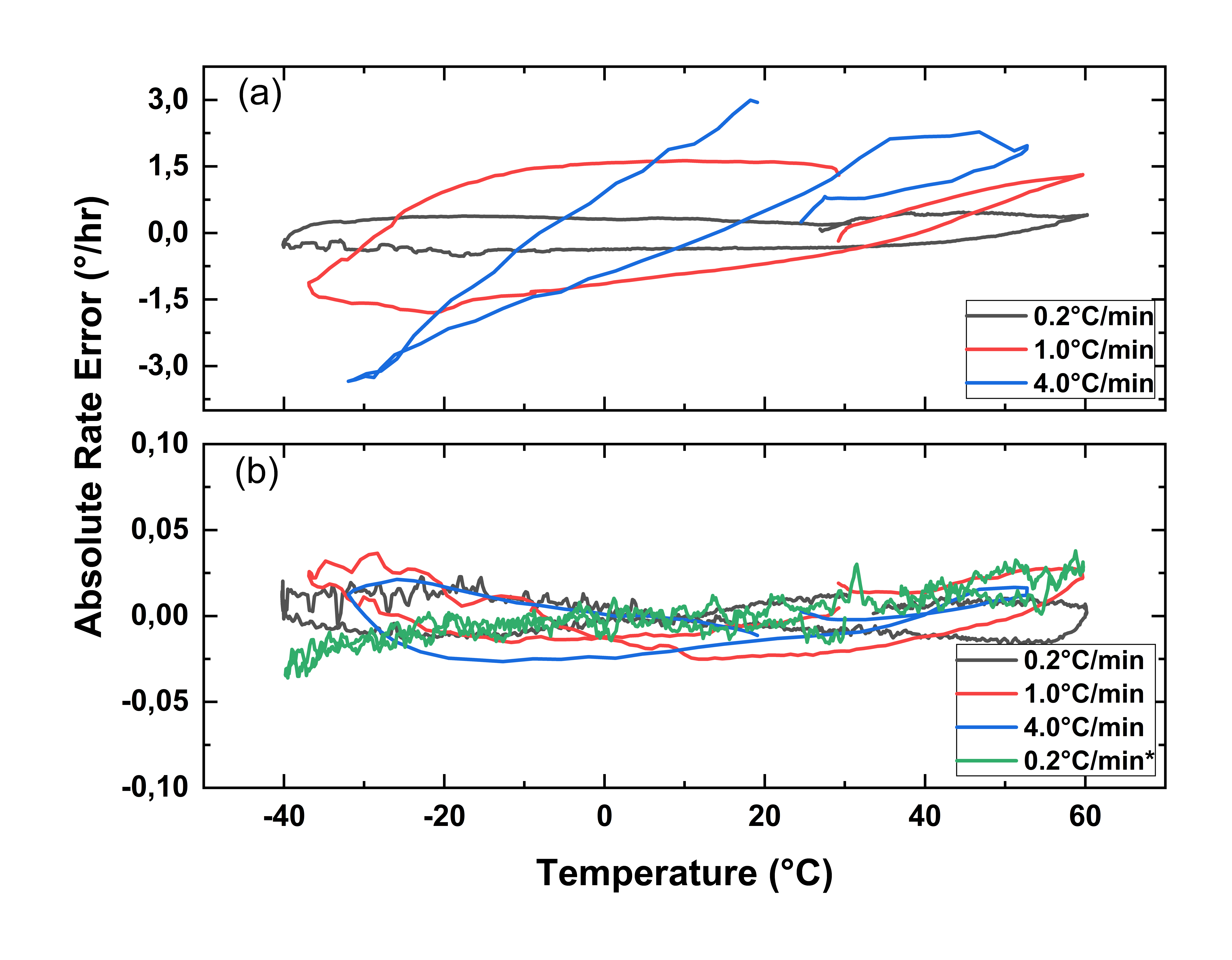}

\caption{\label{fig:Figure_4} (a) Absolute rate errors of the IFOG comprising the quadrupolar fiber coil obtained at a rate of 0.2\degree C/min., 1\degree C/min., and 4\degree C/min. temperature change, (b) compensated absolute rate errors of the quadrupolar fiber coil, respectively, compared with the result obtained from the IFOG comprising a perfectly trimmed hexadecapolar fiber coil at a rate of a 0.2\degree C/min. temperature change.}
\end{figure}

If the real rotation $\phi_{rotation}$ is known, the ratio $R_{C1,C2}$ between $\phi_{shupe_1} $ and $\phi_{shupe_2}$ can be easily found by using Eqs. (\ref{eqnarray:7}), (\ref{eqnarray:8}), (\ref{eqnarray:9}). Here, two different IFOGs comprising hexadecapolar fiber coils were tested at 0.2\degree C/min., 1\degree C/min., and 4\degree C/min. rates of temperature change and the ratio $R_{C1,C2}$ is found to be 3.015, 3.003, and 3.024, respectively for each rate of temperature change, under the Earth rotation. This proves that the ratio of $R_{C1,C2}$ is constant for any rate of temperature change, even if two different fiber coils are used. The uncompensated absolute rate errors are presented for one of the two different hexadecapolar fiber coils at 0.2\degree C/min., 1\degree C/min., and 4\degree C/min rates of temperature in Figure~\ref{fig:Figure_3}(a). The error rate increases with respect to the rate of temperature accordingly. However, the absolute rate error can be compensated down from $\pm2^\circ/$h (as max. rate error) to $\pm$$0.05^\circ/$h as shown in Figure~\ref{fig:Figure_3}(b) by using Eqs. (\ref{eqnarray:9}), and (\ref{eqnarray:10}). The result obtained from the IFOG comprising the perfectly trimmed hexadecapolar fiber coil at a rate of a temperature change of 0.2\degree C/min. is also shown in Figure~\ref{fig:Figure_3}(b), and labeled as 0.2\degree C/min.*, for comparing the results. The data obtained from the reverse axis are provided in the supplementary material (Supp. Figure 3(b)).

To prove the validity of the claim about the elimination of the symmetry dependence of the fiber coils, we used one fiber coil with a hexadecapole winding pattern from the previous setup and one different fiber coil wound with a quadrupole winding pattern that is known to be more sensitive to the Shupe effect. The ratios $R_{C1,C2}$ have also been calculated for these fiber coils and they were found to be 5.167, 5.203, and 5.189, for a 0.2\degree C/min., 1\degree C/min., and 4\degree C/min. rate of temperature change, respectively. Figure~\ref{fig:Figure_4}(a) shows the uncompensated absolute rate error for the quadrupolar fiber coil at 0.2\degree C/min., 1\degree C/min., and 4\degree C/min. rate of temperature change. The absolute rate error can be decreased from $\pm3^\circ/$h (as max. rate error) to $\pm$$0.05^\circ/$h after the compensation as shown in Figure~\ref{fig:Figure_4}(b). The result obtained from the IFOG comprising the perfectly trimmed hexadecapolar fiber coil at a rate of a temperature change of 0.2\degree C/min. is also shown in Figure~\ref{fig:Figure_4}(b), and labeled as 0.2\degree C/min.*, here for comparing the results. The data obtained from the reverse axis are provided in the supplementary material (Supp. Figure 4(b)).

\begin{figure}[!b]
\centering
\includegraphics[width=0.8\textwidth]{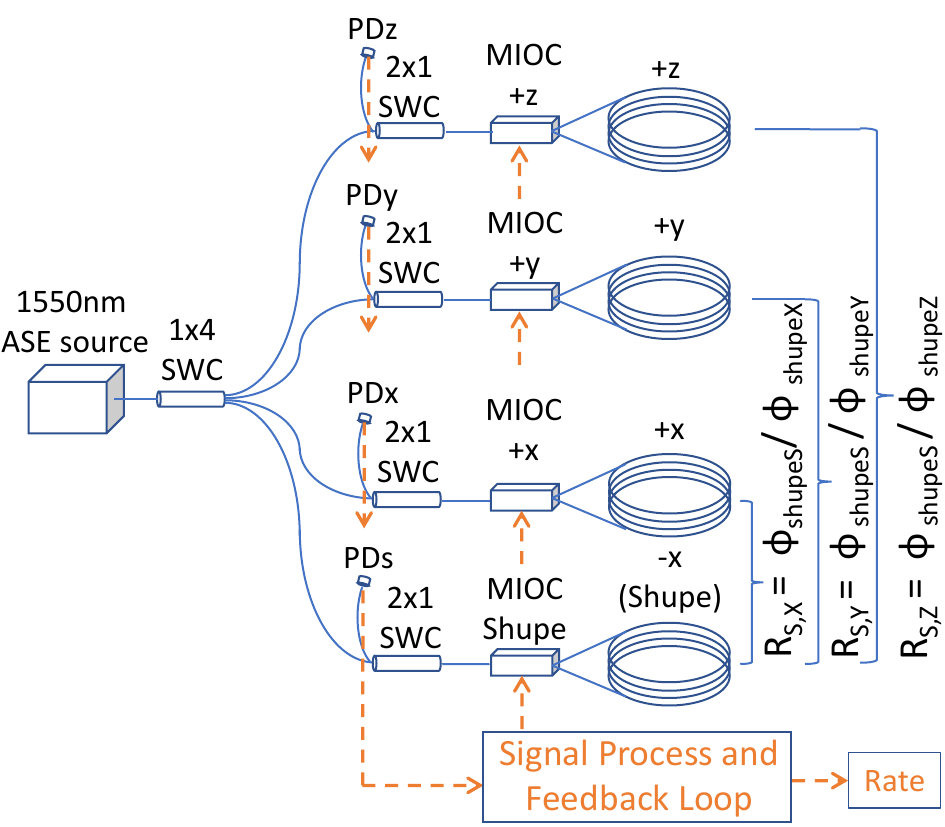}

\caption{\label{fig:Figure_5} Schematics of the configuration of the symmetry-independent-winding IFOG (SIW-IFOG).}
\end{figure}

Based on the agreement that we observed between the real and the compensated results, a new configuration that consists of four-axes, namely -x (as Shupe), +x, +y, and +z, is designed and a novel system called symmetry-independent-winding IFOG (SIW-IFOG) is proposed to compensate for the errors caused by the Shupe effect in Figure~\ref{fig:Figure_5}. The ratios for each different fiber coil pair $R_{S,X}$, $R_{S,Y}$, and $R_{S,Z}$ can be calculated by using Eqs. (\ref{eqnarray:7}), (\ref{eqnarray:8}), and (\ref{eqnarray:9}). Then, we have a constant ratio between the real rate errors coming from the Shupe effect for each different fiber coil pair, such as: 

\begin{eqnarray}
\centering
\label{eqnarray:11} R_{S,i} = \phi_{shupeS}/\phi_{shupe(i)}
\end{eqnarray}
where $i$ represents the axes as X, Y, and Z according to the position of the gyroscope and the ratio $R_{S,i}$ is independent of the rate of temperature change as shown in Figure~\ref{fig:Figure_3} and ~\ref{fig:Figure_4}. 

In the system, the ratio $R_{S,X}$ between CoilS (-x axis) and CoilX (+x axis) will be the checkpoint to calculate how much the real rate errors are coming from the Shupe effect. By adding these two total rates, -x (Eqn. (\ref{eqnarray:13})) and +x (Eqn. (\ref{eqnarray:14})), the real rotations will cancel each other out and the total real rate errors will emerge, as in Eqn. (\ref{eqnarray:15}). The known ratio $R_{S,X}$ reveals the real rate errors for the +x and -x axes. In addition, when the real rate error is known for CoilS, by using the ratio for remaining axes, $R_{S,Y}$, and $R_{S,Z}$, the real rate errors can also be calculated for the y and z axes easily.
\begin{eqnarray}
&& \label{eqnarray:12} R_{S,X} = \phi_{shupeS}/\phi_{shupeX} \rightarrow known\\
&& \label{eqnarray:13}\phi_{total}^{-x} =  \phi_{rotation_{-x}} + \phi_{shupeS}\\
&& \label{eqnarray:14}\phi_{total}^{+x} =  \phi_{rotation_{+x}} + \phi_{shupeX}\\
&& \label{eqnarray:15}\phi_{total}^{-x}+\phi_{total}^{+x} =  \phi_{shupeS} + \phi_{shupeX}
\end{eqnarray}

\section{Conclusion}
In conclusion, starting from the studies aiming to understand the thermal effects that cause phase differences other than the Sagnac effect on different fiber coils, we experimentally showed the independence of the Shupe effect on the real rotation by simply repeating the test with a single hexadecapolar fiber coil in reverse directions. We experimentally calculated a ratio between the real rate errors of the two IFOGs comprising different hexadecapolar fiber coils with the help of the tests conducted at a different rate of temperature changes and showed that this ratio is constant for any rate of temperature change. We also reinforced our argument by repeating the test with a quadrupolar fiber coil that is known to be more sensitive to the Shupe effect in order to prove the claim about the elimination of the symmetry dependence of the fiber coil for the Shupe effect. Based on the outcomes of these studies, we proposed a novel configured complete system called SIW-IFOG that eliminates the winding symmetry dependence of the performance of the IFOGs and compensates the errors 40 times reduced for the hexadecapolar fiber coil and 60 times reduced for the quadrupolar fiber coil down to $\pm$$0.05^\circ/$h that is caused by the Shupe effect. In future studies, the effectiveness of this new method will be examined for rate errors due to the Kerr effect as well as the Faraday effect under temperature change and vibration, which are independent of the direction of the real rotations.

\section*{Acknowledgments}
This work is partially produced from the Ph.D. studies of T.A.S. One of the authors (E.O.) also acknowledges partial support from the Turkish Academy of Sciences.

\bibliographystyle{unsrt}
\bibliography{references}

\begin{thebibliography}{10}

\bibitem{sagnac1913ether}
Georges Sagnac.
\newblock L'{\'e}ther lumineux d{\'e}montr{\'e} par l'effet du vent relatif d'{\'e}ther dans un interf{\'e}rom{\`e}tre en rotation uniforme.
\newblock {\em CR Acad. Sci.}, 157:708--710, 1913.

\bibitem{lefevre2022fiber}
Herve~C Lefevre.
\newblock {\em The fiber-optic gyroscope}.
\newblock Artech house, 2022.

\bibitem{lefevre2014fiber}
Herv{\'e}~C Lef{\`e}vre.
\newblock The fiber-optic gyroscope, a century after sagnac's experiment: The ultimate rotation-sensing technology?
\newblock {\em Comptes Rendus Physique}, 15(10):851--858, 2014.

\bibitem{nayak2011fiber}
Jagannath Nayak.
\newblock Fiber-optic gyroscopes: from design to production.
\newblock {\em Applied Optics}, 50(25):E152--E161, 2011.

\bibitem{yin2017fiber}
Shizhuo Yin, Paul~B Ruffin, and TS~Francis.
\newblock {\em Fiber optic sensors}.
\newblock CRC press, 2017.

\bibitem{lefevre2013fiber}
Herv{\'e}~C Lef{\`e}vre.
\newblock The fiber-optic gyroscope: Challenges to become the ultimate rotation-sensing technology.
\newblock {\em Optical Fiber Technology}, 19(6):828--832, 2013.

\bibitem{korkishko2012interferometric}
Yuri~N Korkishko, Vyacheslav~A Fedorov, Victor Prilutskii, Vladimir~G Ponomarev, Ivan~V Morev, and Sergey~M Kostritskii.
\newblock Interferometric closed-loop fiber-optic gyroscopes.
\newblock In {\em Third Asia Pacific Optical Sensors Conference}, volume 8351, pages 810--817. SPIE, 2012.

\bibitem{lefere1997fundamentals}
Herv{\'e}~C LEF{\`E}RE.
\newblock Fundamentals of the interferometric fiber-optic gyroscope.
\newblock {\em Optical review}, 4(1A):20--27, 1997.

\bibitem{lefevre2012fiber}
Herve~C Lefevre.
\newblock The fiber-optic gyroscope: actually better than the ring-laser gyroscope?
\newblock In {\em OFS2012 22nd International Conference on Optical Fiber Sensors}, volume 8421, pages 63--70. SPIE, 2012.

\bibitem{shupe1980thermally}
David~M Shupe.
\newblock Thermally induced nonreciprocity in the fiber-optic interferometer.
\newblock {\em Applied optics}, 19(5):654--655, 1980.

\bibitem{bohm1982sensitivity}
K~B{\"o}hm, Klaus Petermann, and Edgar Weidel.
\newblock Sensitivity of a fiber-optic gyroscope to environmental magnetic fields.
\newblock {\em Optics Letters}, 7(4):180--182, 1982.

\bibitem{yertutanol2021fiber}
Aylin Yertutanol, {\"O}nder Ak{\c{c}}aalan, Serdar {\"O}{\u{g}}{\"u}t, Ekmel {\"O}zbay, and Abdullah Ceylan.
\newblock Fiber-optic gyroscope for the suppression of a faraday-effect-induced bias error.
\newblock {\em Optics Letters}, 46(17):4328--4331, 2021.

\bibitem{ohno1992development}
Aritaka Ohno, Shinji Motohara, Ryuji Usui, Yuji Itoh, and Kenichi Okada.
\newblock Development of fiber optic gyroscope with environmental ruggedness.
\newblock In {\em Fiber Optic Gyros: 15th Anniversary Conf}, volume 1585, pages 82--88. SPIE, 1992.

\bibitem{frigo1983compensation}
Nicholas~J Frigo.
\newblock Compensation of linear sources of non-reciprocity in sagnac interferometers.
\newblock In {\em Fiber Optic and Laser Sensors I}, volume 412, pages 268--271. SPIE, 1983.

\bibitem{mile2009octupole}
S~Mile.
\newblock Octupole winding pattern for a fiber optic coil.
\newblock {\em European Patent: 2075535A2}, 2009.

\bibitem{chomat1993efficient}
Miroslav Chom{\'a}t.
\newblock Efficient suppression of thermally induced nonreciprocity in fiber-optic sagnac interferometers with novel double-layer winding.
\newblock {\em Applied optics}, 32(13):2289--2291, 1993.

\bibitem{wang2018design}
Yueze Wang, Xiaole Wu, Bohan Liu, and Jun Ma.
\newblock Design and fabrication of high precision optical fiber coil based on temperature error model.
\newblock In {\em Optical Precision Manufacturing, Testing, and Applications}, volume 10847, pages 127--132. SPIE, 2018.

\bibitem{hong2023accurate}
Wei Hong, Xudong Hu, Zerun Zang, Pei Zhang, Shaofeng Lou, Bo~Huang, Yunjiao Li, and Mianzhi Zhang.
\newblock Accurate measurement and enhancement of fiber coil symmetry.
\newblock {\em Applied Optics}, 62(16):E109--E118, 2023.

\bibitem{zhang2020method}
Yunhao Zhang, Zhongxing Gao, Yonggang Zhang, and Liu Yang.
\newblock Method of reducing thermal-induced errors of a fiber optic gyroscope by adding additional winding layers.
\newblock {\em Applied Optics}, 59(8):2462--2467, 2020.

\bibitem{he2019stability}
Dong He, Yangjun Wu, Yulin Li, Zhenrong Zhang, Chao Peng, and Zhengbin Li.
\newblock Stability improvement enabled by four-state modulation in dual-polarization fiber optic gyroscopes.
\newblock {\em Optics Communications}, 452:68--73, 2019.

\bibitem{chen2023suppression}
Yanjun Chen, Yuwen Cao, Lanxin Zhu, Yan He, Wenbo Wang, Huimin Huang, Xiangdong Ma, and Zhengbin Li.
\newblock Suppression of effects of temperature variation by high-order frequency modulation in large fiber-optic gyroscopes.
\newblock {\em Applied Physics Letters}, 122(14), 2023.

\bibitem{song2023advanced}
Ningfang Song, Xiaobin Xu, Zuchen Zhang, Fuyu Gao, and Xiaowei Wang.
\newblock Advanced interferometric fiber optic gyroscope for inertial sensing: A review.
\newblock {\em Journal of Lightwave Technology}, 2023.

\end{thebibliography}


\begin{thebibliography}{1}

\bibitem{shupe1980thermally}
David~M Shupe.
\newblock Thermally induced nonreciprocity in the fiber-optic interferometer.
\newblock {\em Applied optics}, 19(5):654--655, 1980.

\bibitem{zhang2017quantitative}
Zhuo Zhang and Fei Yu.
\newblock Quantitative analysis for the effect of the thermal physical property parameter of adhesive on the thermal performance of the quadrupolar fiber coil.
\newblock {\em Optics express}, 25(24):30513--30525, 2017.

\bibitem{goettsche1996trimming}
Randy~P Goettsche and Ralph~A Bergh.
\newblock Trimming of fiber optic winding and method of achieving same, June~18 1996.
\newblock US Patent 5,528,715.

\bibitem{li2013novel}
Zhihong Li, Zhuo Meng, Tiegen Liu, and X~Steve Yao.
\newblock A novel method for determining and improving the quality of a quadrupolar fiber gyro coil under temperature variations.
\newblock {\em Optics Express}, 21(2):2521--2530, 2013.

\bibitem{albert1983computer}
Mary~Remley Albert and Gary~E Phetteplace.
\newblock Computer models for two-dimensional steady-state heat conduction.
\newblock {\em Computer models for two-dimensional steady-state heat conduction}, 1983.

\end{thebibliography}

\end{document}


\section{Supplementary Material}

We studied the thermal characteristics of fiber coils with different patterns theoretically and experimentally. We built up a numerical model and showed that the use of this winding method will help us to obtain the best thermal performance in an interferometric fiber optic gyroscope (IFOG) that consists of the respective fiber coil. We used the trimming method in addition to the usage of this special winding method for the very same purpose. We also verified the model experimentally with different rates of temperature change. 

If a fiber coil experiences a temperature change over time, thermally triggered non-reciprocity arises because the two counter-propagating beams will pass through the same region of the coil at different times. The underlying reason for this effect is that when the fiber is exposed to thermal changes, the light propagation constant ($\beta$) in the fiber changes at different rates along the fiber.

Beams traveling uneven paths create undesirable nonreciprocal phase shifts along with the real phase shift coming from the rotation that is to be measured and by considering this phase shift, the angular error of thermally induced non-reciprocity for the IFOG can be calculated by \cite{shupe1980thermally}

\begin{eqnarray}
\phi_{shupe} (t) =&& \frac{n_c}{4NA}(\frac{dn_c}{dT}+\alpha n_c)(\int_{0}^{L/2} dl(2l-L)\nonumber\\
&&\times
[\Delta T (l,t) - \Delta T (l',t)]),
\end{eqnarray}

where $n_c = 1.46$ is the refractive index of the fiber core, $N = 48$ is the number of turns in the coil, $A$ is the area of the coil with a diameter of $10$ cm, $dn_c/dT = 10^{-5} / ^{\circ}C$ is the temperature dependence of the refractive index of the fiber, $\alpha = 5x10^{-7} / ^{\circ}C$ is the coefficient of linear thermal expansion of fiberglass and $L = 1037 m$ is the total length of the coil. Here, $l$ and $l'$ represent the locations of the two rotating beams at the same time and $\Delta T(l,t)$, and $\Delta T(l',t)$ represent the temperature differences between the locations $l$ and $l'$ simultaneously.

After the simplification for a coil which is symmetrically wound from its midpoint, the equation becomes

\begin{eqnarray}
\phi_{shupe}  (t) =&&\gamma \sum_{i=1}^{M/2} [\sum_{j=1}^{S}[\Delta T (A_{i,j},t)-\Delta T (B_{i,j},t)] \nonumber\\
  &&\times\int_{(i-1)L/M/S}^{iL/M/S} dl(2l-L)],\\
&& \gamma  = \frac{n_c}{4NA} (\frac{dn_c}{dT} + \alpha n_c)
\end{eqnarray}

where $(A_{i,j},t)$ and $(B_{i,j},t)$ represent the clockwise and counterclockwise beam positions on the coil at the same time, $M = 48$ is the number of layers, and $S$ is the number of $dl$ section on each layer. This formula examines the temperature variations for each layer and each $dl$ length corresponding to that layer. In the present paper, since the 2D conduction thermal model is used for temperature distribution on the coil, $S$ is taken as 80 and that also represents the number of turns for each layer of the fiber coil that is used in experiments.

\begin{figure}[H]
\centering
\includegraphics[width=0.65\textwidth]{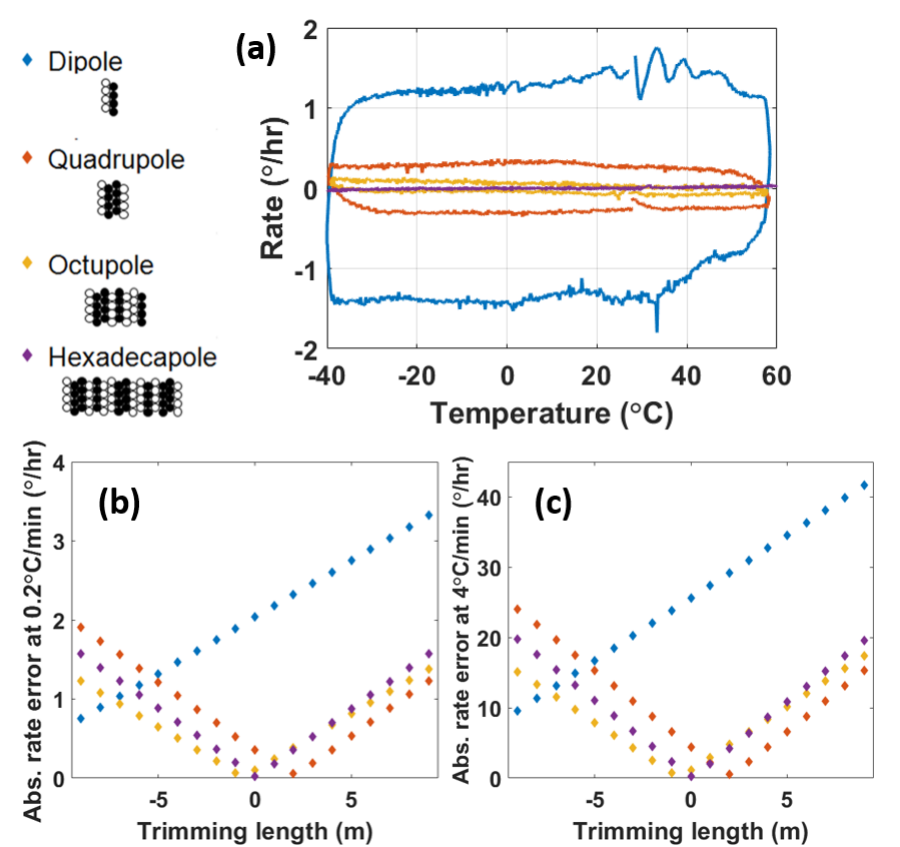}
\caption{(a) Schematics of the winding methods of 4 different fiber coils and a comparison of the corresponding experimental rates of temperature change of 0.2\degree C/min. Numerical results for the absolute rate error vs. the trimming length of the fiber coils with different winding methods at a (b) 0.2\degree C/min. and (c) 4\degree C/min. rate of temperature change.}
\label{fig:Supp.Figure_1.pdf} 
\end{figure}

\begin{figure} [H]
\centering
\begin{subfigure}{1\textwidth}
\setlength{\belowcaptionskip}{0pt}
  \centering
  \includegraphics[width=.75\linewidth]{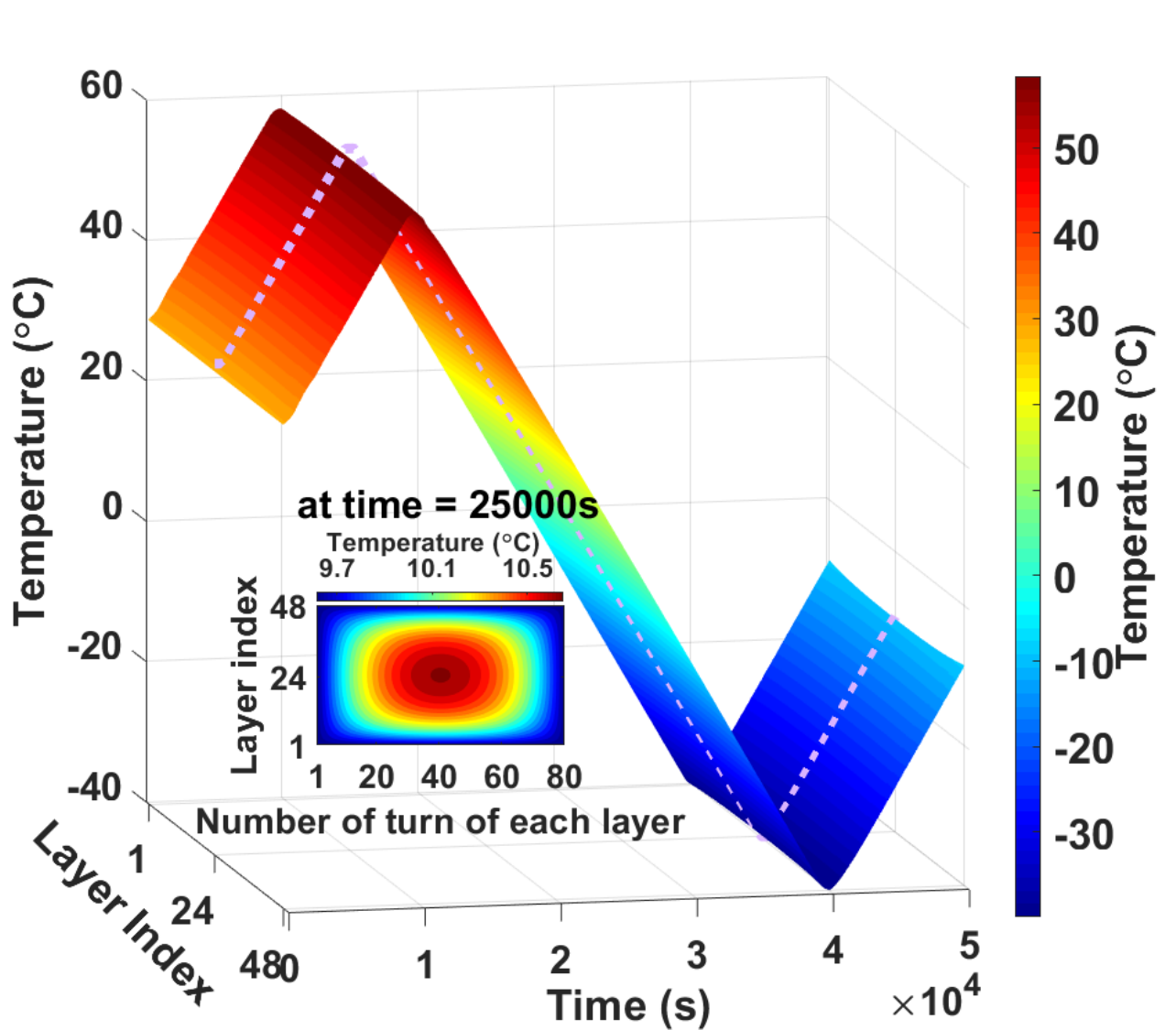}
  \caption{Rate of temperature change 0.2\degree C/min.}
  \label{fig:Supp.Figure_2a}
\end{subfigure}%

\begin{subfigure}{1\textwidth}
\setlength{\belowcaptionskip}{0pt}
  \centering
  \includegraphics[width=.75\linewidth]{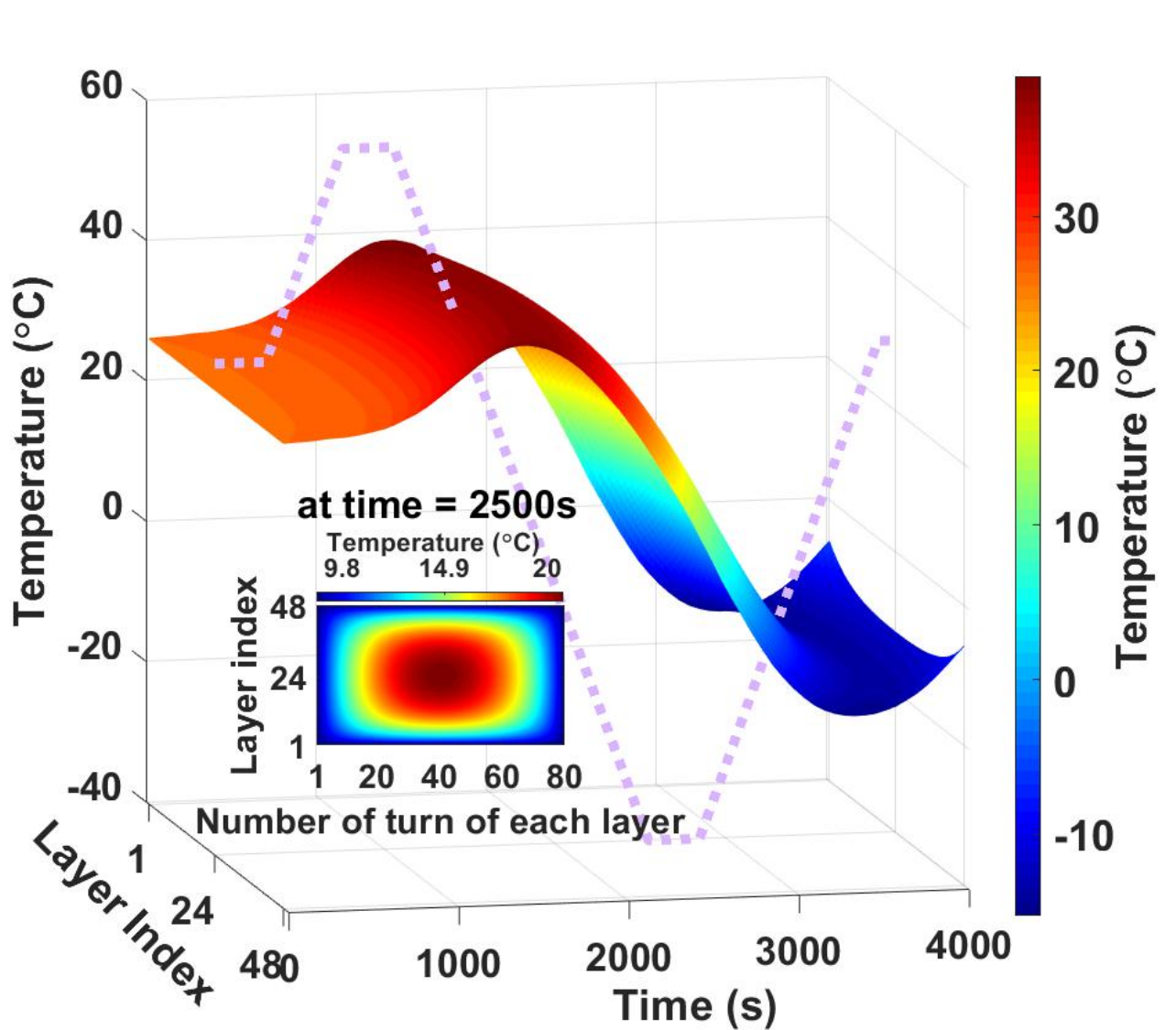}
  \caption{Rate of temperature change 4\degree C/min.}
  \label{fig:Supp.Figure_2b}
\end{subfigure}
\caption{Temperature distribution curves on each layer of the fiber coil with respect to time at different rates of temperature change and heat transfers of the layers with respect to the number of turns of each layer at certain times.}
\label{fig:Supp.Figure_2}
\end{figure}

Figure~\ref{fig:Supp.Figure_1.pdf} (a) shows the geometric illustrations of the portions of repetitive sections of the fiber coils with different winding patterns along with the corresponding experimental results obtained from the IFOG comprising these fiber coils at a 0.2\degree C/min. rate of temperature change and Figure~\ref{fig:Supp.Figure_1.pdf} (b-c) shows the trimming needs obtained from the calculations at 0.2\degree C/min. and 4\degree C/min. rates of temperature change. In Figure~\ref{fig:Supp.Figure_1.pdf} (a) without trimming, 1.755, 0.356, 0.126, and 0.021\degree /h absolute rate errors were obtained from dipole (AB), quadrupole (ABBA), octupole (ABBABAAB) and hexadecapole (ABBABAABBAABABBA) winding patterns, respectively, at a 0.2\degree C/min. rate of temperature change, experimentally. Moreover, in Figure~\ref{fig:Supp.Figure_1.pdf} (b) 2.043, 0.353, 0.100, and 0.020\degree /h absolute rate errors were also obtained numerically computing the results obtained from each pattern based on the previous formula showing a good agreement with the experiments. It was observed that the hexadecapole pattern outperforms the others in terms of the need for trimming as it presents approximately 100, 15, and 5  times reduced absolute rate error than dipole, quadrupole, and octupole pattern when tested under a 0.2\degree C/min. rate of temperature change for both simulations and experiments. As shown in Figure~\ref{fig:Supp.Figure_1.pdf} (b-c), the need of trimming decreases as the symmetry of the winding pattern increases at both rates of temperature change. The graphs of the rate of temperature changes inside the coil for 0.2\degree C/min. and 4\degree C/min. were also given in Figure~\ref{fig:Supp.Figure_2} with dashed lines.

Although it was attempted to keep the winding method as symmetrical as possible, the nature of the fiber winding has a catch to it. Inevitable errors that come with the process itself such as the stress applied to the fiber during the winding, the change of the fiber length with temperature, the impurities, inhomogeneities, or the change of refractive index along the fiber \cite{zhang2017quantitative} are always there. That is why, even if it is sought to produce a symmetrically wound fiber optic coil, the thermal performance of the fiber coil may still need to be improved at the end. Trimming is offered here as an auxiliary method to overcome this problem by basically shifting the midpoint of the fiber coil by adding or trimming a certain portion of the fiber ends \cite{goettsche1996trimming}. In such a case, trimming shifts the midpoint of the fiber coil as $l_{eff} / 2$ \cite{li2013novel}. Then, the total angular error of thermally induced nonreciprocity including the asymmetry coming from this shift of the coil's midpoint can be calculated as

\begin{eqnarray}
\phi_{shupe} (t) =&&\gamma \Bigg\{ \sum_{i=1}^{M/2} [\sum_{j=1}^{S}[\Delta T (A_{i,j},t)-\Delta T (B_{i,j},t)] \nonumber\\ 
&&\times\int_{(i-1)L/M/S+\frac{I_{eff}}{2}}^{iL/M/S-\frac{I_{eff}}{2}} dl(2l-L)] \nonumber\\
&&+ \sum_{i=1}^{M/2-1} [\sum_{j=1}^{S}[\Delta T (A_{i,j},t)-\Delta T (B_{i,j},t)] \nonumber\\
&&\times\int_{iL/M/S-\frac{I_{eff}}{2}}^{iL/M/S+\frac{I_{eff}}{2}} dl(2l-L)]  \Bigg\}
\end{eqnarray}

The heat transfer method is used for a fixed coil geometry, material, and temperature profile to find $\Delta T $. We use the explicit model of unsteady-state equation discretization by using a constant thermal diffusivity $\alpha = 5x10^{-9} m^2 / s$ for fiberglass to calculate the temperature change in time \cite{albert1983computer}.

\begin{eqnarray}
 \frac{\delta T}{\delta t} = \alpha (\frac{\delta^2 T}{\delta x^2} + \frac{\delta^2 T}{\delta y^2})
\end{eqnarray}

 After the discretization of the equation, it becomes
 
 \begin{eqnarray}
 \frac{T_M^{n+1} - T_M^n}{\Delta t} = \alpha (\frac{T_L - 2T_M + T_R}{\delta x^2} + \frac{T_T - 2T_M + T_B}{\delta y^2})
\end{eqnarray}

\begin{equation}
\left\{
 T_M^{n+1} =  T_M^n + k_1 (T_L - 2T_M + T_R)^n+ k_2 (T_T - 2T_M + T_B)^n
 \right\}
\end{equation}

where $ k_1 = \alpha \Delta t/\Delta x^2 $ and $k_2 = \alpha \Delta t/\Delta y^2$,  $T_L$, $T_T$, $T_M$, $T_B$, $T_R$ represent left, top, middle, bottom, and right boundary respectively if the conductivity $k_{1,2}$ were to be constant over the region.

Figure ~\ref{fig:Supp.Figure_2} shows the temperature distributions on each layer of the fiber coil with respect to the time and layer index at different rates of temperature change. Dashed lines indicate the set temperature of the climatic chamber where the fiber coil is tested. The cross-sections of the layers and turns of the fiber coil at certain times are also given as a function of temperature. While the rate of temperature change is low, e.g., 0.2\degree C/min., the set temperature and the temperature felt by the coil almost match and the temperature difference between the coldest and the hottest point of the fiber coil becomes 1.1\degree C at t=25000s. However, when we increase the rate of the temperature change up to 4\degree C/min., the set temperature is no longer felt by the fiber coil properly and the temperature difference between the coldest and the hottest point of the fiber coil becomes 10.5\degree C at t=2500s. The temperature variation itself among the layers also varies upon different rates of temperature change.

\begin{figure}[H]
    \centering
    \includegraphics[width=1\textwidth]{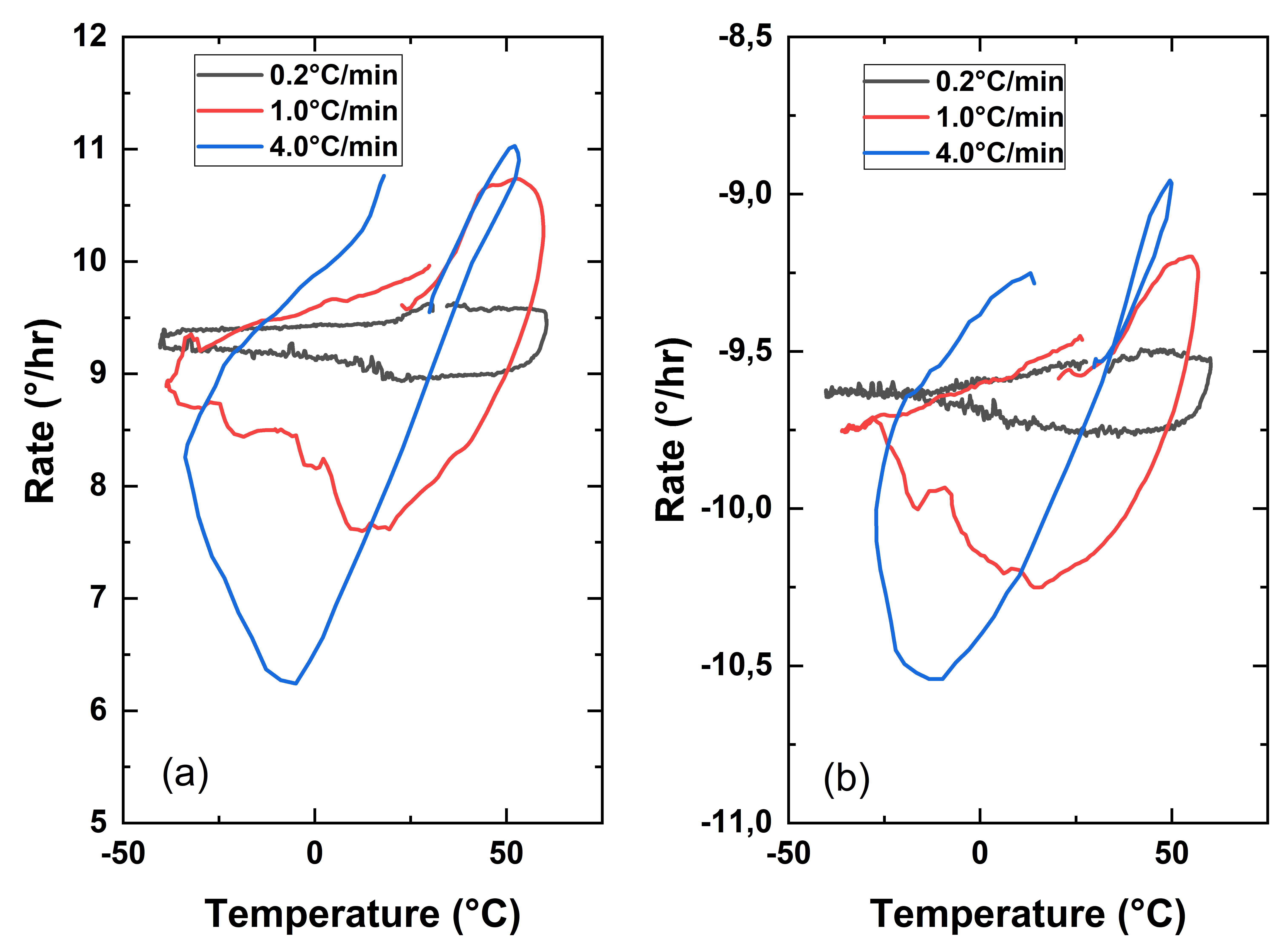}
    \caption{Rate vs. temperature graphs of the IFOGs comprising (a) the horizontally placed hexadecapolar fiber coil, (b) horizontally reversed placed hexadecapolar fiber coil at a rate of 0.2\degree C/min., 1\degree C/min., and 4\degree C/min. temperature change}
    \label{fig:Supp.Figure_3.png}
\end{figure}

\begin{figure}[H]
    \centering
    \includegraphics[width=1\textwidth]{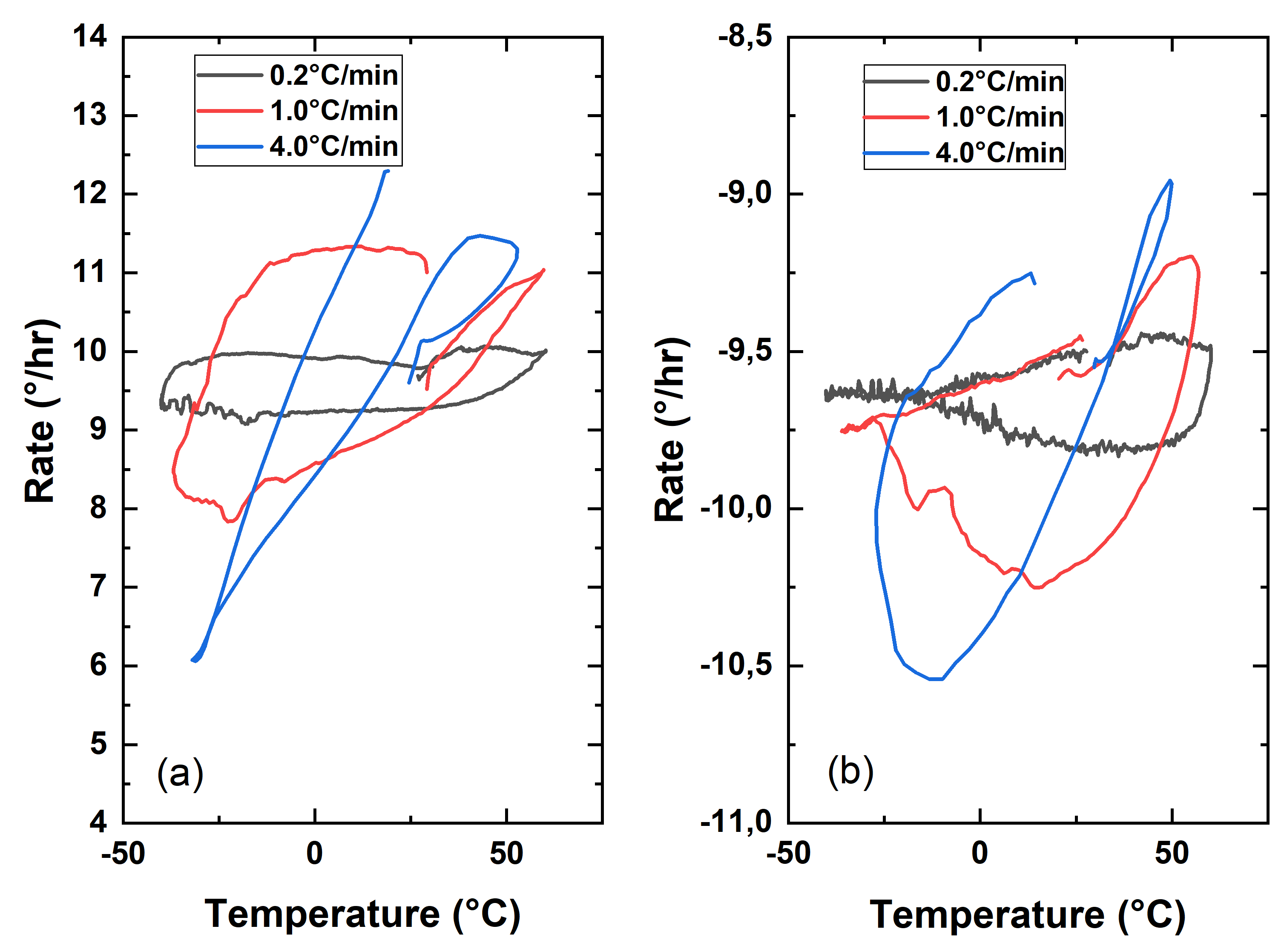}
    \caption{Rate vs. temperature graphs of the IFOGs comprising (a) the horizontally placed quadrupolar fiber coil, (b) horizontally reversed placed hexadecapolar fiber coil at a rate of 0.2\degree C/min., 1\degree C/min., and 4\degree C/min. temperature change}
    \label{fig:Supp.Figure_4.png}
\end{figure}

We observed that the Shupe effect is independent of the direction of the real rotation as shown in Figure 1 in the main article. We kept running the tests by using pairs of hexadecapolar and quadrupolar fiber coils at a different rate of temperature changes in order to further investigate the thermal effects on different fiber coils as shown in Figure ~\ref{fig:Supp.Figure_3.png} and Figure ~\ref{fig:Supp.Figure_4.png}. We used the experimental setup of which details are described in the main article.

\begin{figure}[H]
\setlength{\belowcaptionskip}{0pt}
    \centering
    \includegraphics[width=1\textwidth]{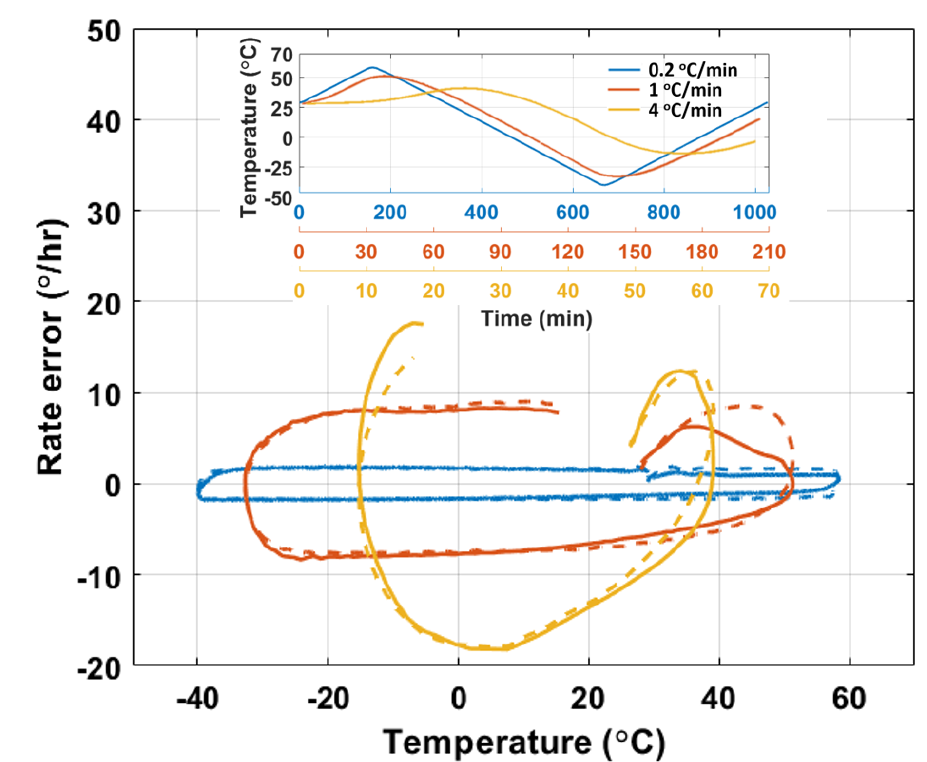}
    
    \caption{Comparison of the rate data (around zero) of the IFOG comprising the fiber coil with the hexadecapole pattern with different rates of temperature change (inset-sensor reading on coil) obtained from experiments (solid lines) and simulations (dashed lines) at a 10 m trimming length.}
    \label{fig:Supp.Figure_5}
\end{figure}

Figure~\ref{fig:Supp.Figure_5} shows the experimental and simulated absolute rate errors at three different rates of temperature change of 0.2, 1, and 4\degree C/min. at 10 m trimming length. 1.704\degree /h absolute rate error was measured at a rate of 0.2\degree C/min. temperature change with both experiments and simulations. 8.239\degree /h and 9.007\degree /h at a rate of 1\degree C/min., 17.99\degree /h and 18.15\degree /h at a rate of 4\degree C/min. temperature change with experiments and simulations, respectively.

\begin{figure}[H]
\setlength{\belowcaptionskip}{0pt}
    \centering
    \includegraphics[width=1\textwidth]{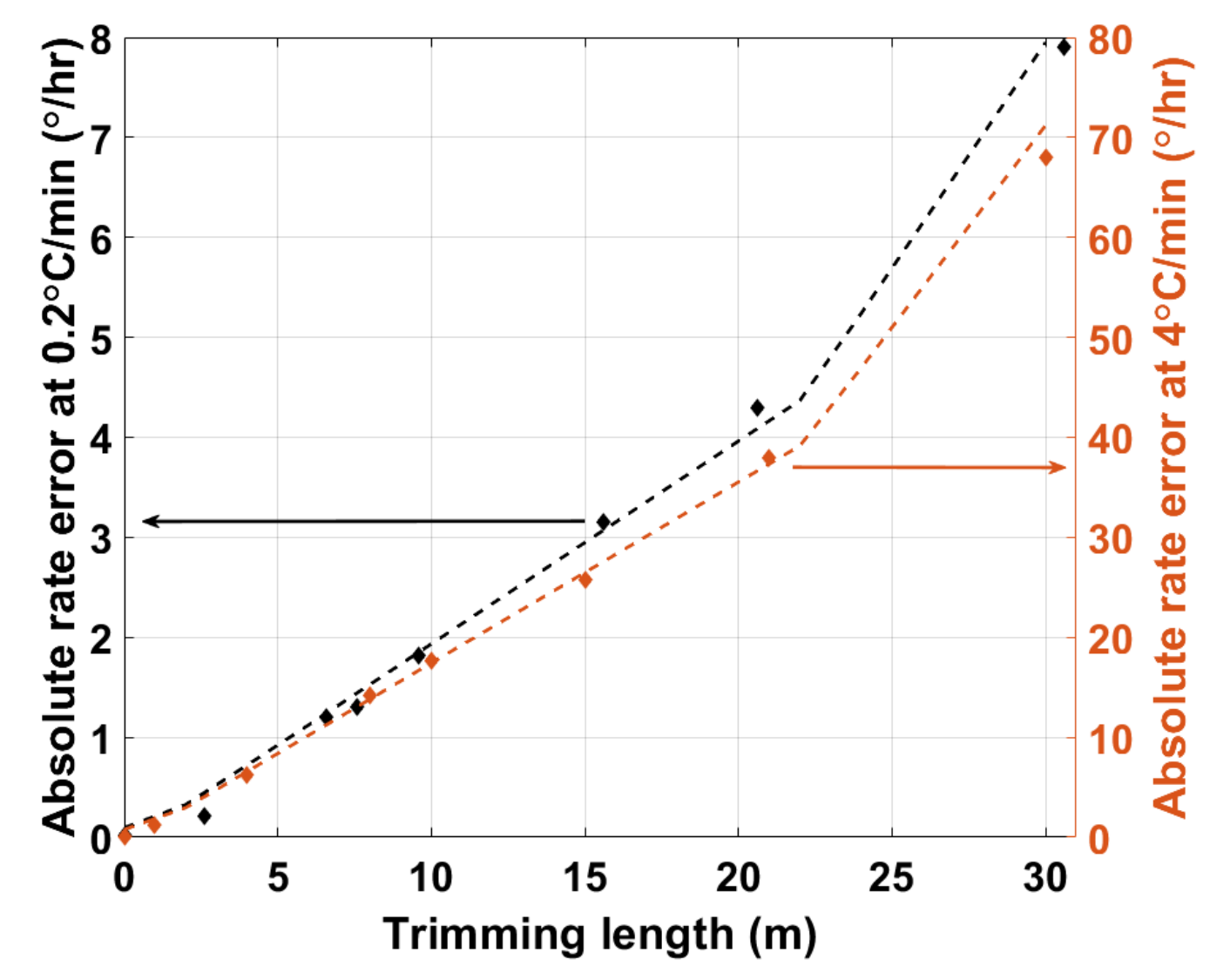}

    \caption{Comparison of the experimental (dots) and simulated (dashes) absolute rate errors at the rates of temperature change of 0.2\degree C/min. and 4\degree C/min. with respect to the trimming length.}
    \label{fig:Supp.Figure_6}
\end{figure}

The agreement of the simulated and experimental results at different rates of temperature changes supports the reliability of the thermal model that we used. Based on this, we investigated the effect of the trimming length of the fiber coil on the absolute rate error. Figure~\ref{fig:Supp.Figure_6} shows the relation between the absolute rate errors with respect to the trimming length of the fiber coil at different rates of temperature change. 0.023\degree /h absolute rate error was obtained with simulations indicated with dashed lines and 0.024\degree /h absolute rate error was obtained experimentally indicated with dots at the rate of temperature change of 0.2\degree C/min. Likewise, 0.023\degree /h absolute rate error was obtained with simulations and 0.035\degree /h absolute rate error was obtained experimentally at a rate of temperature change of 4\degree C/min. showing a good agreement with each other. In order to obtain these results, we still need to find and adjust the perfect trimming amount on the fiber coil at the order of 10 cm length. There, the trimming length is directly proportional to the absolute rate error. Therefore, increasing the trimming length in such coils worsens the thermal performance of the IFOG comprising the corresponding fiber coil as it increases the absolute rate error. In addition to this, as shown in Figure~\ref{fig:Supp.Figure_6}, there is a clear slope difference between shorter and longer trimming lengths at about 20 m and that is because the trimming length is longer than about 20 m that is longer than one layer of the coil that is used in the setup, causing a change in the symmetry of the hexadecapole pattern.

\bibliographystyle{unsrt}
\bibliography{supplement}